\documentclass[sigconf]{acmart}

\usepackage{booktabs} 

\usepackage{siunitx}
\usepackage{multirow}
\usepackage{adjustbox}
\usepackage{color}
\usepackage[utf8]{inputenc}
\usepackage[T1]{fontenc}
\usepackage{xspace}
\usepackage{tabularx}
\usepackage{graphicx}
\usepackage{bm}
\usepackage{enumitem}
\usepackage{balance}

\usepackage{pifont}

\DeclareMathOperator*{\argmin}{arg\,min}

\makeatletter
\DeclareRobustCommand\onedot{\futurelet\@let@token\@onedot}
\def\@onedot{\ifx\@let@token.\else.\null\fi\xspace}

\def\eg{\emph{e.g}\onedot} 
\def\ie{\emph{i.e}\onedot}

\def\wrt{w.r.t\onedot} 

\makeatother

\setcopyright{acmlicensed}

\theoremstyle{definition}
\newtheorem{definition}{Definition}

\begin{document}

\copyrightyear{2019}
\acmYear{2019}
\acmConference[MM '19]{Proceedings of the 27th ACM International Conference on Multimedia}{October 21--25, 2019}{Nice, France}
\acmBooktitle{Proceedings of the 27th ACM International Conference on Multimedia (MM '19), October 21--25, 2019, Nice, France}
\acmPrice{15.00}
\acmDOI{10.1145/3343031.3351036}
\acmISBN{978-1-4503-6889-6/19/10}

\fancyhead{}

\title{Diachronic Cross-modal Embeddings~\textsuperscript{*}}

\thanks{* Please cite the ACM MM 2019 version of this paper.}

\author{David Semedo}
\affiliation{%
  \institution{NOVALINCS}
  \streetaddress{}
  \city{Universidade NOVA de Lisboa}
  \state{Portugal}
}
\email{df.semedo@campus.fct.unl.pt}

\author{João Magalhães}
\affiliation{%
  \institution{NOVALINCS}
  \streetaddress{}
  \city{Universidade NOVA de Lisboa}
  \country{Portugal}}
\email{jm.magalhaes@fct.unl.pt}

\renewcommand{\shortauthors}{D. Semedo et al.}

\begin{abstract}
Understanding the semantic shifts of multimodal information is only possible with models that capture cross-modal interactions over time.
Under this paradigm, a new embedding is needed that structures visual-textual interactions according to the temporal dimension, thus, preserving data's original temporal organisation.
This paper introduces a novel \textit{diachronic cross-modal embedding} (DCM), where cross-modal correlations are represented in embedding space, throughout the temporal dimension, preserving semantic similarity at each instant $t$.
To achieve this, we trained a neural cross-modal architecture, under a novel ranking loss strategy, that for each multimodal instance, enforces neighbour instances' temporal alignment, through subspace structuring constraints based on a temporal alignment window.
Experimental results show that our DCM embedding successfully organises instances over time. Quantitative experiments, confirm that DCM is able to preserve semantic cross-modal correlations at each instant $t$ while also providing better alignment capabilities. Qualitative experiments unveil new ways to browse multimodal content and hint that multimodal understanding tasks can benefit from this new embedding.
\end{abstract}

%
%



\maketitle

\graphicspath{{figs/}}

\section{Introduction}

\begin{figure*}[t]
    \centering
     \includegraphics[width=0.9\linewidth]{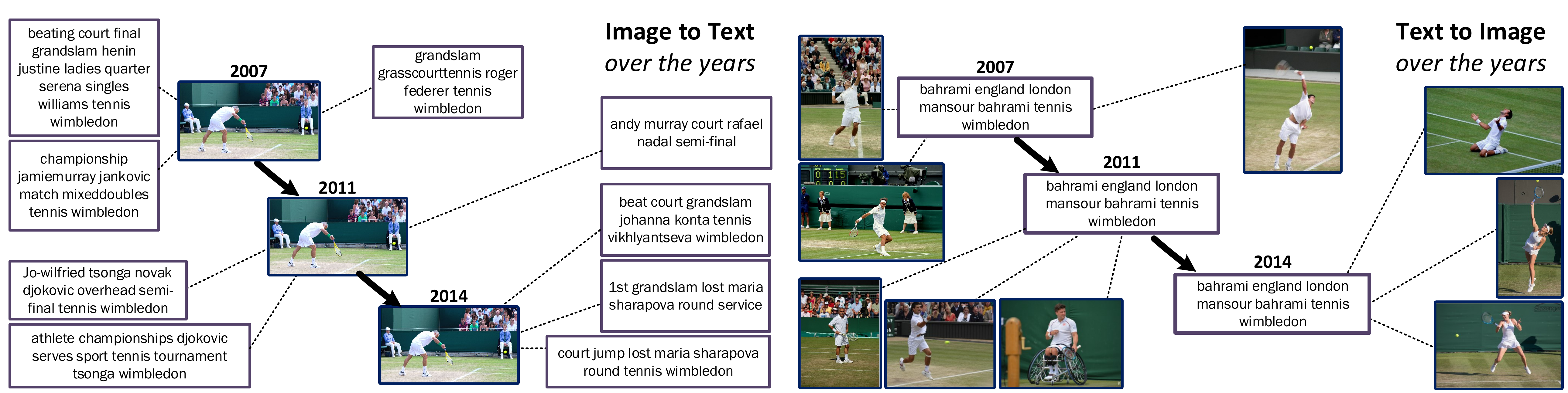}
\caption{Diachronic Cross-modal Embeddings illustration.}
  \label{fig:temporal_embedding_illustration}
\end{figure*}

It is known that with just a few glances, humans can quickly and accurately understand an image with minimal effort~\cite{FeiFei2007WhatDW}.
One of the reasons for this phenomenom is that images, or more generally humans' visual perception of elements of the physical world (\eg objects, shapes), does not change over time.
Language on the other hand, is a way for humans to express/communicate their knowledge about something. Accordingly, textual descriptions of images reflect the way humans refer to each image at a \textit{given point in time}, thus being susceptible to change over time, either due to the occurrence of external events~\cite{C12-1093} or simply by word meaning change across time~\cite{hamilton2016diachronic,Yao:2018:DWE:3159652.3159703}. Recently, word embedding models that capture language evolution over time (referred as  \textit{distributional diachronic} models) have been proposed~\cite{Yao:2018:DWE:3159652.3159703,N18-1044,hamilton2016diachronic}. In our setting we seek to obtain representations that \textit{capture the evolution of visual and textual correlations over time}. While visual elements can be seen as anchors -- \ie pictures and videos freeze a reality and do not change afterwards -- word descriptions change over time (\eg the same image, or a semantically similar one, may be referred twice, at different points in time, but with different descriptions).

Cross-modal embeddings, structure images and texts in a common space, that is organised based on data interactions, enabling the study of such interactions towards multimedia understanding. Until now, these embeddings have been learned in a \textit{static} manner, \ie without preserving the time dimension, and thus ignoring the evolution of modality interactions~\cite{Rasiwasia:2010:NAC:1873951.1873987, Feng:2014:CRC:2647868.2654902, 7298966,Fan:2017:CRL:3123266.3123369,8013822,Wang:2017:ACR:3123266.3123326,7346492,Xu:2018:MSL:3206025.3206033,DBLP:journals/corr/abs-1708-04308,Semedo:2018:TCR:3240508.3240665,Peng:2016:CSR:3061053.3061157}.
Approaches have ranged from solutions that organize the space according to linear correlations~\cite{Rasiwasia:2010:NAC:1873951.1873987, 7298966, 7780910} (image and texts co-occurrence), semantic~\cite{Wu:2018:LSS:3240508.3240521,Wang:2017:ACR:3123266.3123326,Peng:2016:CSR:3061053.3061157,Xu:2018:MSL:3206025.3206033} (category information) and/or temporal correlations~\cite{Semedo:2018:TCR:3240508.3240665}.
State-of-the-art methods can preserve semantic and temporal correlations but do not consider the time dimension in its original scale, thus resulting in the loss of all the information regarding the evolution of data.

In this paper we depart from previous approaches and propose a novel model for learning \textit{Diachronic Cross-modal embeddings}, materialised by continuous projection functions, where the time dimension is preserved in order to capture data interactions over time.
Figure~\ref{fig:temporal_embedding_illustration} illustrates the target cross-modal diachronic embedding. The hypothesis is that cross-modal interactions evolve along the temporal dimension. Therefore, the embedding space should structure images and texts such that for each instant $t$, i.e., elements are organised according to semantic correlations between instances and corresponding absolute timestamp.
This results in a model in which neighbours of an element (e.g. a text or an image), at time instant $t_1$, may differ from the neighbours at time instant $t_2$, if data correlations between the two instants change (Figure~\ref{fig:temporal_embedding_illustration}).

The first challenge we need to address corresponds to unveiling and quantifying, for each image and text, the evolution of the semantic correlations \wrt to other instances.
The second, corresponds to the learning of the diachronic embedding, in which at each instant $t$, semantic category information is used to guide the structuring of multimodal instances' neighbourhood. This requires solving the \textit{alignment problem}, in which embeddings of an instance at an instant $t$, should retain correlation with embeddings of semantically similar instances, on distinct instants.
We address these challenges, by employing a two-part approach for neighbourhood structuring for an arbitrary instant time $t$: first, for instances of the same semantic category within a given time range, semantic correlations need to be maximal, second, instances outside the time range are placed far apart.
Then, a novel ranking loss is formulated to achieve a continuous diachronic structure, by enforcing correlations from the two dimensions in the space structure organisation, by using temporal context of instances, at each instant, while aligning instances across adjacent time instants.

In summary, the key contributions of this paper are:
\begin{enumerate}
    \item The first Diachronic Cross-modal embedding learning approach, where the evolution of multimodal data correlations are modelled. Time is explicitly modelled, thus allowing \textit{conditioning on time at both training and inference time};
    \item A novel  \emph{temporally constrained ranking loss} formulation, aligns instances embeddings over time, and enables the learning of neural projections from timestamped multimodal data;
    \item A principled approach that offers statistical guarantees, and allows for correct joint-inferences (image+text+time) that other methods do not, enabling it to be used for a wide number of media interpretation tasks.
\end{enumerate}

\section{Related Work}
Across the literature, several works have researched methods to model and incorporate temporal aspects to better understand data. There is enough evidence in the literature that demonstrates how information sources are correlated along a timeline of events, with different media and event types~\cite{Kim:2013:TWI:2433396.2433417,C12-1093,Tsytsarau:2014:DNE:2623330.2623670,Sakaki:2010:EST:1772690.1772777, 10.1007/978-3-642-41181-6_73}.
In~\cite{Kim:2013:TWI:2433396.2433417} temporal clues were used to improve search relevance at query time, by modelling content streams using a multi-task regression on multivariate point processes.
In~\citet{10.1007/978-3-642-41181-6_73}, the value of temporal information for the tasks of image annotation and retrieval, such as tag frequency, is recognised.

In order to model the temporal behaviour of data, embeddings must retain temporal correlations~\cite{Kulkarni:2015:SSD:2736277.2741627,hamilton2016diachronic,pmlr-v70-bamler17a,Yao:2018:DWE:3159652.3159703,N18-1044, Semedo:2018:TCR:3240508.3240665}. The challenge resides in capturing such correlations and incorporating them in cross-modal embeddings.
~\citet{Blei:2006:DTM:1143844.1143859} proposed a dynamic topic model to capture temporal behaviour of data by modelling the evolution of word interactions over time. The model is an extension of static topic models (\eg LDA), where latent topic evolution over time was accounted.
Topic model approaches treat words as \emph{symbols} and thus lack all the properties of distributed representations~\cite{Bengio:2003:NPL:944919.944966}. Word embedding models aim at learning word representations, such that words that appear in similar contexts are structured close together in the embedding space~\cite{DBLP:journals/corr/abs-1301-3781}. \textit{Diachronic Word Embeddings} consist of word embeddings that model word meaning change across time, by encoding words' usage over time~\cite{Kulkarni:2015:SSD:2736277.2741627,hamilton2016diachronic,pmlr-v70-bamler17a,Yao:2018:DWE:3159652.3159703,N18-1044}. Lately these models have been actively researched to aid the understanding of words' semantic evolution. A common approach to learn such embeddings has been to split text documents into bins (\eg by year), and then train a static Skip-Gram~\cite{DBLP:journals/corr/abs-1301-3781} (\emph{word2vec}) model on each bin. Embeddings of adjacent bins are then aligned by learning a linear transformation that performs the best rotational alignment, while preserving cosine similarities~\cite{Kulkarni:2015:SSD:2736277.2741627,hamilton2016diachronic,Yao:2018:DWE:3159652.3159703}. Data binning introduces several issues and limitations. Small bins are required to capture fine-grained interactions, however these may incur in bins with very few data for training. Conversely, with large bins only coarse grained representations can be obtained. To overcome this,~\citet{N18-1044} recently proposed a continuous approach, in which time is taken as a continuous variable. The model learns an embedding for each word $w$ at each time instant $t$. Our work goes in this direction, however two aspects invalidate the use of existing word diachronic models:  a) unlike words, that are predominant across time instants, each instance is posted only once, invalidating existing alignment strategies,  b) in the cross-modal scenario two modalities need to be aligned instead of only one.

Static cross-modal embedding models represent multimodal data in  a common space. Early approaches~\cite{Rasiwasia:2010:NAC:1873951.1873987,Gong:2014:MES:2584252.2584265,7006724,6587747}, learn projections based on linear correlation. State-of-the-art works adopt neural networks, to learn non-linear modality projection functions~\cite{Andrew:2013:DCC:3042817.3043076,Feng:2014:CRC:2647868.2654902,7298966,Peng:2016:CSR:3061053.3061157,Wang:2017:ACR:3123266.3123326,8013822}. The focus has been on capturing correlations between instances (using category information when available), without accounting for temporal correlations, for data organisation. Recently, a temporal cross-modal common space approach was proposed~\cite{Semedo:2018:TCR:3240508.3240665}, where temporal correlations are used to organise instances in a static embedding space. The authors observed that for dynamic data, the incorporation of temporal insights increases retrieval performance. This embedding space has the limitation that it is not diachronic, since the temporal dimension is discarded after training, thus not allowing the study of the evolution of semantic correlations. We depart from their work by proposing the first diachronic cross-modal embedding, in which semantic evolution is preserved.

\begin{figure*}[t]
\centering
  \includegraphics[width=1.0\linewidth]{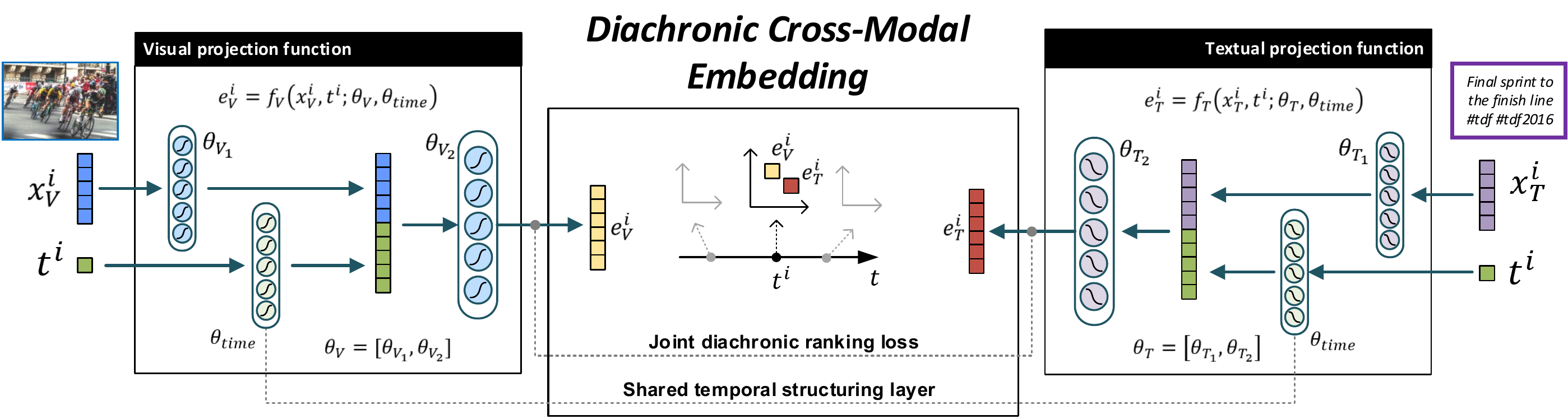}
  \caption{Diachronic cross-modal architecture overview.  Visual (blue) and textual (purple) instances, at an instant $t^i$, are mapped to a $D$ dimensional diachronic embedding space. A shared temporal structuring layer takes the timestamp $t^i$ as input and learns an embedding for $t^i$, that is then used to independently condition modality projections on time. A diachronic ranking loss is responsible for structuring instances over time. Best viewed in color.}
    \label{fig:framework_overview}
\end{figure*}
\section{Diachronic Cross-Modal Embedding}
The key element of diachronic embedding methods is the set of time-preserving projection functions, responsible for mapping the original data onto the embedding. This section will detail the proposed embedding and the corresponding project
in functions.
\subsection{Definitions}
\label{subsec:definitions}
We start by introducing some notation and defining the task of diachronic cross-modal embedding learning. Without loss of generality, let  $\mathcal{C} =\{d_i\}^N_{i=1}$  be a set of $N$ \emph{visual-textual} instance tuples
\begin{equation}
d^i=(\vec{x}_V^i, \vec{x}_T^i, ts^i, c^i),
\end{equation}
where $\vec{x}_V^i \in \mathcal{R}^{D_V}$ and $\vec{x}_T^i \in \mathcal{R}^{D_T}$ are the feature representations of the image and textual elements, respectively, $ts^i$ the timestamp and $c^i$ the instance (unique) semantic category. Accordingly, $D_V$ and $D_T$ correspond to the image and text features dimensionality, respectively.
The instances timestamp have a timespan defined by $TS = [t_s, t_f]$, where $t_s$ and $t_f$ are the first and last instants of the dataset, respectively. Let $*\in\{V,T\}$ on the remainder of this paper, to avoid notation cluttering.

The goal is to obtain a common continuous (over time instants) embedding space, in which the visual and textual elements are organised according to their semantic category and timestamp. The space is formally defined as follows:

\begin{definition}{A \emph{diachronic cross-modal embedding space}}
refers to a common space, that structures the visual and textual elements of the data instances over time. In this embedding, similarity between instances of the same category that are close in time, is maximised. In all other cases, similarity between instances is minimal.
\end{definition}

In diachronic word embeddings, the temporal dimension captures \emph{word meaning change}~\cite{hamilton2016diachronic,Yao:2018:DWE:3159652.3159703}. In a cross-modal scenario, with timestamped data,
the words (textual modality) that co-occur with pictures (visual modality), at different instants in time, may be different. This characteristic is quantified by the semantic alignment of the diachronic models:

\begin{definition}{The \textit{semantic alignment over time}} is the embeddings correlation, between semantically similar instances, that is retained across different time instants.
\end{definition}

\subsection{Time-preserving Projections}
\label{subsec:formalisation}
The diachronic cross-modal embedding embeds the modality vector $\vec{x}^i_*$ (image or text) of an instance $d^i$ and a time instant $t$, using the pair of functions:
\begin{equation}
\vec{e}_V=f_V( \vec{x}^i_V, t; \bm{\theta}_V,\bm{ \theta}_{time})\hspace{5mm}\vec{e}_T=f_T( \vec{x}^i_T, t;\bm{ \theta}_T,\bm{ \theta}_{time}),
\end{equation}
where each $\vec{e}_*\in \mathbb{R}^D$ denotes the embedding of the corresponding modality, at time instant $t$. $\bm{\theta}_V = [\bm{\theta}_{V_h};\bm{\theta}_{V_o}]$ and $\theta_T = [\bm{\theta}_{T_h};\bm{\theta}_{T_o}]$ are the model parameters, and $\bm{\theta}_{time}$ are the key parameters responsible for controlling the temporal structuring of the $\vec{x}^i_V$ and $\vec{x}^i_T$ projections. As a consequence, parameters $\bm{\theta}_{time}$ tie the two projection functions.

Diachronic cross-modal embedding functions are defined by the mappings:
\begin{equation}
f_V(\cdot): \mathbb{R}^{D_V}\times TS \mapsto \mathbb{R}^D \hspace{1cm} f_T(\cdot): \mathbb{R}^{D_T}\times TS \mapsto \mathbb{R}^D.
\end{equation}
The output of $f_V$ and $f_T$ is normalised such that $\ell_2(f_{*}(\cdot)) = 1$. Accordingly, instances will be organised based on time and semantic similarity, over a $D$-dimensional space.
Similarity between projected sample elements $x^i_*$ and $x^j_*$, is computed through \emph{cosine} similarity.

The diachronic projection functions $f_V(\cdot)$ and $f_T(\cdot)$ are implemented as a neural network with 2 fully connected layers.
Figure~\ref{fig:framework_overview} depicts the neural architecture.
Formally, the diachronic projection functions are defined as
\begin{equation}
    f_{*}(x_*^i,ts^i)=tanh\left(  \bm{\theta}_{*_o}\cdot \big[f_{h_*}(x_*^i);\  f_{time}(ts^i)\big]\ \right),
\end{equation}
\begin{equation}
    f_{*_h}(x_*^i)=tanh(\bm{\theta}_{*_h}\cdot x_*^i), \qquad f_{time}(ts^i)=tanh(\bm{\theta}_{time}\cdot ts^i),
\end{equation}
where $\bm{\theta}_{*_h}$, $\bm{\theta}_{time}$ and $\bm{\theta}_{*_o}$ correspond to hidden (per modality), time and output layer weight matrices, respectively. $[\cdot]$ denotes the concatenation operation.
An initial \textbf{encoding layer} ($f_{V_h}$ or $f_{T_h}$), receives the input vector and transforms it into an internal representation that is compatible
with the internal representation of data timestamps. A \textbf{shared time embedding layer} ($f_{time}$) maps data timestamps to an embedding representation. The obtained time embedding is then used to condition the output projections of $f_{V_h}$ and $f_{T_h}$, through a concatenation operation, making them time-dependent. A final \textbf{output layer} takes as input the result of conditioning  $f_{*_h}$ and $f_{time}$ to produce the final $D$-dimensional projection to a diachronical  embedding space.

The resulting embeddings, produced by $f_V$ and $f_T$ are characterised by the properties defined in the folllowing section.

\subsection{Embedding Properties}
\label{subsec:properties}
The structure of the temporal embedding space, \ie how multimodal instances will be organised,  is formalised by a set of fundamental properties.
These properties stem from two grounding intuitions: data is primarily associated by the time dimension, and then by their semantic categories.
The model will thus capture the evolution of semantic correlations, over time instants, by maximising the similarity between instances that are within a given temporal window and share the same category.
Accordingly, two embedding vectors $\vec{e}^i_*$ and $\vec{e}^j_*$ will be projected into the same neighbourhood if the two following properties are met simultaneously:
\begin{itemize}
    \item \textbf{Property 1.} the timestamps of $d^i$ and $d^j$ are within the same temporal window, \ie $|t^i-t^j| < w$, and the two instances $d^i$ and $d^j$ share the same category, \ie  $c^i = c^j$.
\end{itemize}
Conversely, the two embedding vectors $\vec{e}^i_*$ and $\vec{e}^j_*$ will be projected onto distant regions if:
\begin{itemize}
    \item \textbf{Property 2.} the timestamps of $d^i$ and $d^j$ are outside the same temporal window, \ie $|t^i-t^j| > w$, independently of the instances' semantic category;
    \item \textbf{Property 3.} elements do not share any semantic category.
\end{itemize}
\vspace{4pt}
The final and most novel property follows from the requirement that the target embedding space needs to be continuous over time. Thus, a final \textit{semantic alignment over time}  property is introduced:
\begin{itemize}
\item \textbf{Property 4.} For each image or text of an instance $d^i$, embeddings must evolve smoothly between neighbouring time instants $t^1$ and $t^2$, with $|t^2 - t^1| \leq w$.
\end{itemize}

\section{Diachronic Embedding Learning}
To learn the time-dependent continuous embedding functions $f_V(\cdot)$ and $f_T(\cdot)$ it is essential to maximise correlation in the new embedding space. In this section we define the optimisation objective and show how it enforces the temporal organisation of the embedding.

\subsection{From Projections to Ranking Loss}
\label{subsec:architecture}
Following the definition of the diachronic projection functions, the two component correlation scheme (temporal and semantic), and respecting the properties defined in section~\ref{subsec:properties}, we start by noting that similarities in the embedding space are computed by the dot product   $s(x^i_*,x^j_*)=f_{*}(x^i_*)\cdot f_{*}(x^j_*)$.
Building on the most recent state-of-the-art cross-modal learning works~\cite{7410369,7780910,Wang:2017:ACR:3123266.3123326,8013822,Semedo:2018:TCR:3240508.3240665}, we adopt the \textit{ranking loss function} as the model base loss. In its general formulation, triplets $(x_*^a, x_*^p, x_*^n)$, are composed by an anchor element $x_*^a$, that should be more similar to positive elements $x_*^p$ sharing a category, than to negative elements $x_*^n$ not sharing categories, by at least a margin $m$.
Triplet constraints are expressed as $s(x_*^a, x_*^p) > s(x_*^a, x_*^n) + m$, and then turned into a differentiable function, by means of a relaxation under the hinge loss function~\cite{herbrich2000large}:
\begin{equation}
    \ell_\theta(x_*^a,x_*^p,x_*^n)=[m - s(x_*^a, x_*^p) + s(x_*^a, x_*^n)]_+,
\label{eq:constraint_hinge}
\end{equation}
where $m$ denotes a constant margin,  $[x]_+$ the function $max(0,x)$, and $\theta = [\bm{\theta}_V; \bm{\theta}_T; \bm{\theta}_{time}]$ is the complete set of parameters. One of such constraints would then be enforced for each sampled triplet. In the next section we detail how ranking loss is extended to cope with the temporal dimension.

\subsection{Joint Diachronic Ranking Loss}
The learning problem is then formulated by coupling the learning of the two individual modality and a third timestamp embedding functions, through a global loss function $\mathcal{L}$.
The full loss function of our model, for diachronic cross-modal embedding learning, is derived by enforcing multiple constraints, for each possible anchor element, and summing all the constraint violations. To this end, we define each $f_*$ as a neural network, capable of unveiling complex non-linear interactions.
The objective function becomes
\begin{equation}
\argmin_{\bm{\theta}_V, \bm{\theta}_T, \bm{\theta}_{time}} \mathcal{L}_{\bm{\theta}_V, \bm{\theta}_T, \bm{\theta}_{time}}(\mathcal{C}),
\label{eq:general_loss}
\end{equation}
with $\bm{\theta}_V$, $\bm{\theta}_T$ and $\bm{\theta}_{time}$ being the projection functions parameters.

State-of-the-art cross-modal retrieval interlace modalities by enforcing triplet constraints in both modality directions~\cite{7780910,Wang:2017:ACR:3123266.3123326, Mithun:2018:LJE:3206025.3206064}, \ie $image\mapsto text$ and $text\mapsto image$.
Thus, we formulate the final loss $\mathcal{L}$ function for temporal cross-modal embedding model with parameters $\theta = [\bm{\theta}_V; \bm{\theta}_T; \bm{\theta}_{time}]$ as:
\begin{equation}
\mathcal{L}_\theta(\mathcal{C}) = \sum_{a,p,n} \underbrace{\mathcal{L}_{\theta}(x_V^a,x_T^p,x_T^n)}_{image\ \mapsto\  text} +  \underbrace{\mathcal{L}_{\theta}(x_T^a,x_V^p,x_V^n)}_{text\ \mapsto\  image},
\label{eq:pairwise_loss}
\end{equation}
where $p$ and $n$ denote indices of positive and negative instances, respectively, \wrt to an anchor element $x_*^a$. This function is evaluated batch-wise. Thus, at each batch, the sampled elements are used to create triplet constraints.

\subsection{Continuous Diachronic Structure}
\label{subsec:continuous_structure}
Formally, for each anchor element $x_*^a$, of a triplet $(x_*^a, x_*^p, x_*^n)$, we define the following loss function:
\begin{equation}
\mathcal{L}_\theta(x_*^a, x_*^p, x_*^n)=\\\mathcal{L}_{inter}(x_*^a, x_*^n) + \mathcal{L}_{intra}(x_*^a, x_*^p),
\label{eq:smoothed_ranking_loss}
\end{equation}
where $\mathcal{L}_{inter}$ and $\mathcal{L}_{intra}$ are based on the ranking loss function, enforcing inter-category  and intra-category embedding related properties, respectively. Both $x_*^n$ and $x_*^p$ correspond to sampled negative and positive images or texts, respectively.

\vspace{3mm}
\noindent
\textbf{Temporal triplets.}
To let the loss function of equation~\ref{eq:smoothed_ranking_loss} enforce the required properties, special attention needs to be given to triplets' temporal and semantic correlations. Given an anchor element $x_*^a$, a temporal window of size $w$ is used to enforce smoothness between temporal neighbourhoods in the embedding, for images and texts of the same category as $x_*^a$. This context-window formulation has been quite successful in word embeddings~\cite{Mikolov:2013:DRW:2999792.2999959, D14-1162}. For diachronic cross-modal embeddings, we consider temporally adjacent multimodal instances instead of a sliding window over text.
Thus, triplets are obtained by sampling positive and negative instances by always using as positive image/text to the anchor $x_*^a$, its modality counterpart of the same instance $a$ (e.g. for the anchor $x_V^a$ we use $x_T^{p=a}$ as positive, and vice-versa).
With this strategy we maximise correlation between modalities using images and texts that occur together, allowing us to better capture intra-category semantic diversity.

\vspace{3mm}
\noindent
\textbf{Intra-category and temporal smoothing.}
The assumption is that given an image or a text, its embedding should change smoothly between adjacent time instants. Smoothness of that change is captured by the size of the considered window (\textit{Property 4}).
The function $\mathcal{L}_{intra}$ is responsible for embedding alignment over time, \ie \textit{Property 4}. To accomplish this, a temporal window of size $w$ is considered. The rationale is that we want that instances of the same category to be close. But from \textit{Property 2} definition, embeddings of instances of the same category, should be far apart, if they are temporally far apart. Thus, if we do not enforce any margin between positive instances, the optimal solution is when all instances of the same category are mapped to the same point, losing temporal evolution of semantic correlations. To overcome this, we employ a temporally decaying ranking loss formulation, for instances of the same category. Namely, given an anchor element $x_*^a$ and a positive sampled instance $x_*^p$,  $\mathcal{L}_{intra}$ is defined as the following branch function:
\begin{equation}
    \mathcal{L}_{intra}(x_*^a,x_*^p) = \left\{
\begin{array}{ll}
      0 &, |t^a - t^p|\leq w \\
      \rho(t^a, t^p)\cdot\ell_{\theta}(x_{V/T}^a, x_{T/V}^{a}, x_*^p) &,\ otherwise\\
\end{array}
\right.
\end{equation}
where $\rho(t^a, t^b)= 1-exp(-|t^a-t^b|\cdot\lambda)$ is a temporal decaying function, and $\lambda$ the decay rate. When two positive instances are less than $w$ instants far part, no margin is enforced between the two. Otherwise, a triplet constraint is enforced, weighted by a decaying function that exponentially decreases the importance of $\ell_{\theta}$, the closer in time two instances are.

Finally, it is important to observe that the triplet sampling for batch creation makes a stochastic approximation to the optimal set of triplets (mining the optimal triplets for each batch is computationally too expensive~\cite{2015arXiv150303832S}). Hence, although for a given batch it is likely that a given anchor $x_*^a$ with timestamp $t^a$, has no adjacent (in time) projections, that same anchor $x_*^a$ will occur with high-likelihood in subsequent batches with the required instances. Thus, convergence guarantees are preserved, due to the stochastic approximation made through the triplet sampling strategy.

\vspace{3mm}
\noindent
\textbf{Inter-category separation.}
For inter-class alignment $\mathcal{L}_{inter}$, we want to structure the embedding space such that instances of different categories will be far apart, independently of their timestamp. To this end, we enforce a triplet loss constraint, with a large margin (\ie $m=1$) over such triplets.
Formally, $\mathcal{L}_{inter}$ is defined as:
\begin{equation}
\mathcal{L}_{inter}(x_*^a, x_*^n)=\ell_{t}(x_*^a, x_*^{(p=a)}, x_*^n),
\label{eq:negatives_window_loss}
\end{equation}
where if the anchor is an image (\ie $x_V^a$), then the positive corresponds to a text (\ie $x_T^a$), and vice-versa. This formulation achieves two goals: enforces the separation of positive from negative instances, by a margin $m$, and aligns the embeddings of the image and the text of the anchor instance.

\subsection{Binned Diachronic Structure}
\label{subsec:binned_diachronic}
The previous general formulation will behave as a binned diachronic method~\cite{hamilton2016diachronic,Kulkarni:2015:SSD:2736277.2741627} when $\mathcal{L}_{intra}(\cdot) = 0$. In this extreme case, data can be first divided into bins and a static cross-modal embedding model is trained on data from each bin. Then, embeddings of adjacent bins are aligned by solving the Orthogonal Procrustes problem~\cite{hamilton2016diachronic, Kulkarni:2015:SSD:2736277.2741627}. Namely, given two embedding matrices $\bm{M}_{t}$ and $ \bm{M}_{t+1}$ with shape $N\times D$, containing $N$ images and texts embeddings representative of adjacent time instants, the best rotational alignment is computed as:
\begin{equation}
    \bm{\Omega}_{t\mapsto t+1} =\argmin_{\bm{\Omega}^T\bm{\Omega}=\bm{I}} \lVert \bm{M}_{t} \bm{\Omega}  -  \bm{M}_{t+1}\rVert_F,
\end{equation}
preserving cosine similarities within each $\bm{M}$.
For diachronic word embeddings, words present at one instant are also present in the next. Thus, one can directly align the embeddings of each word, at different instants. As in the cross-modal scenario each document occurs only once, we set $\bm{M}_{t+1}$ by projecting images and texts from instant $t$ in bin $t+1$, and perform alignment.
This approach, while being prone to embedding alignment issues, is taken as a strong baseline. It preserves temporal locality biases by definition, as cross-modal correlations over distinct bins are never considered.

\section{Evaluation}

\subsection{A 20 years Flickr Images Dataset}
\begin{figure}[t]
    \centering
     \includegraphics[width=0.6
     \linewidth]{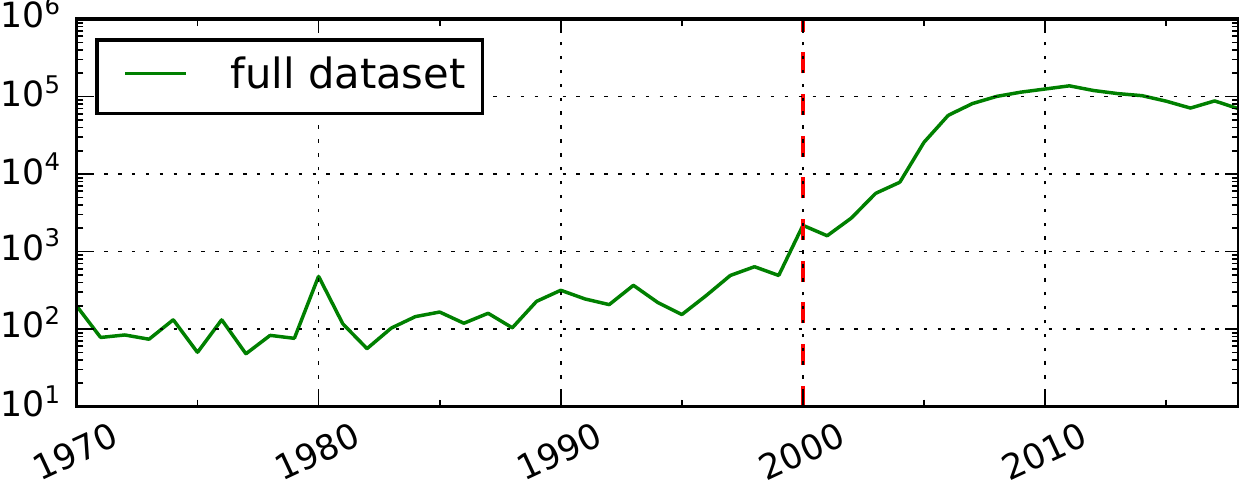}
     \includegraphics[width=0.49\linewidth]{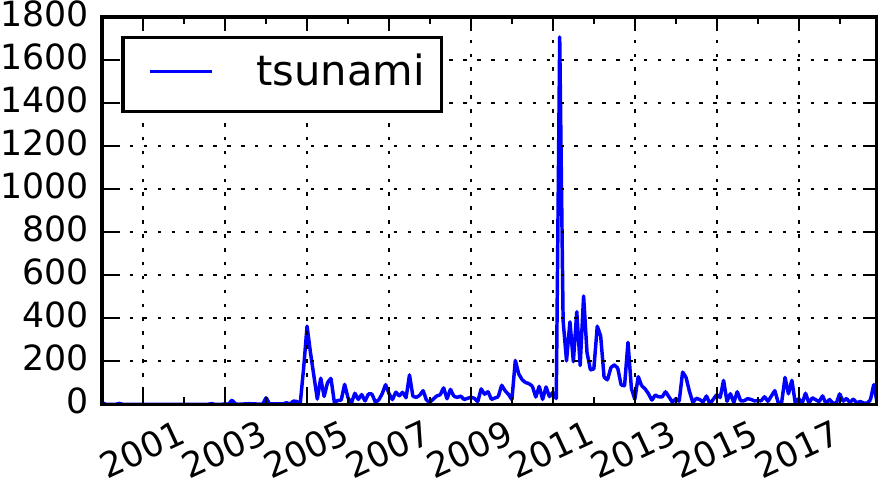}
     \includegraphics[width=0.49\linewidth]{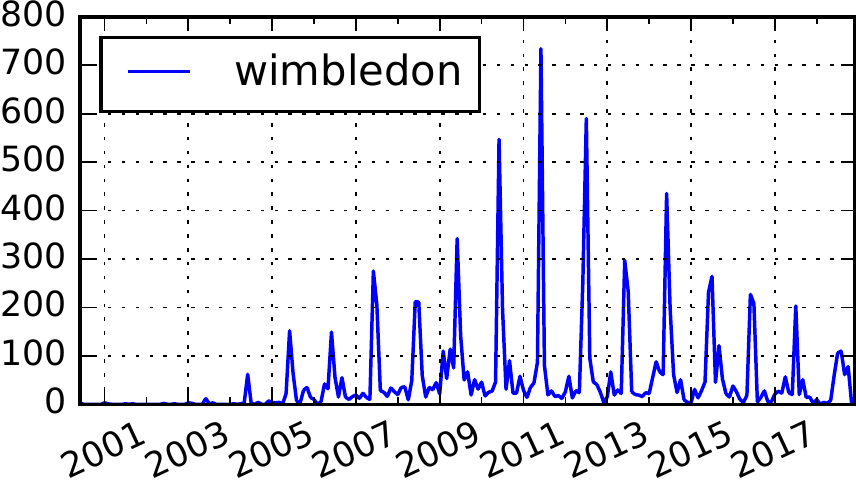}
    \caption{Temporal distribution of documents across the full dataset and four sample categories.}
    \vspace{-15pt}
  \label{fig:category_temporal_freq}
\end{figure}

For evaluation we construct a new large scale weakly-labeled dataset\footnote{\url{https://novasearch.org/multimodal-diachronic-models}} with multimodal instances obtained from the Flickr\footnote{\url{https://www.flickr.com/}} social network. Namely, we collect documents related with topics that show a dynamic behaviour over time such as spike-based and recurrent events. Figure~\ref{fig:category_temporal_freq} shows the temporal distributions of four sampled categories, and illustrates the diversity in terms of dynamic behaviour captured by the dataset.  Data is collected over the period 1-1-1970 to 31-12-2018. The Flickr API is used\footnotemark ~to retrieve images and texts from a total of 21 categories: \textit{easter-sunday}, \textit{edinburgh-festival}, \textit{flood}, \textit{formula-one}, \textit{horse-riding}, \textit{independence-day},  \textit{london-marathon}, \textit{mountain-camping}, \textit{nuclear-disaster}, \textit{olympic-games}, \textit{picnic}, \textit{rock-climbing}, \textit{scuba-diving}, \textit{snowboarding}, \textit{solar-eclipse}, \textit{terrorism}, \textit{tour-de-france}, \textit{tsunami}, \textit{white-house}, \textit{wimbledon}, \textit{world-cup}. We use the category name as keyword to query the API and collect data, and filter instances whose \textit{date taken} is outside the considered temporal range. The models' granularity is set to \emph{months}. To ensure that enough instances are available for each bin, we restrict the temporal range of images to the past 20 years (red line on figure~\ref{fig:category_temporal_freq} depicts the cut), and bins with less than 100 documents are excluded. After applying a set of SPAM filtering techniques, we obtain a total of 709033 instances. In general, images have (near) professional quality. Texts are on average 23.0 words long. We use 10\% of the data for testing and split the remaining data in 90\% for training and 10\% for validation, resulting in 574,308, 63,804 and 70,921 instances, for training, validation and testing, respectively.

\footnotetext{Only Creative Commons licensed data was retrieved.}

\subsection{Implementation Details}
Networks are jointly trained using SGD, with $0.9$ momentum, and a learning rate $\eta=0.005$. We train the model for $25$ epochs and retain the best performing model based on the validation set loss. Mini-batch size is set to $64$. For each neuron, we use \emph{tanh} non-linearities.
A pre-trained ResNet-50~\cite{He_2016_CVPR}, with the last fully connected layer removed (softmax), is used for image representation. We set $\lambda=0.1$, window size $w=4$ and pairwise-ranking loss margin $m=1.0$. We adopt the TF-IDF bag-of-words representation for texts and CNN image representations for all methods.

The layers corresponding to the $\bm{\theta}_{*_h}$ and $\bm{\theta}_{time}$ parameters have  dimension 1024 and 200 respectively, and $\bm{\theta}_{*_o}$ has $D=200$ dimensions. Thus, for an instance $d^i$, the visual projection network takes the CNN representation $x_V^i$ of the image, the textual projection a bag-of-words representation of the text $x_T^i$, and the timestamp embedding the timestamp as input, producing the $D$-dimensional diachronic embedding.

\subsection{Methodology}
Experiments are performed in a cross-modal setting with different temporal parameters, \ie when an image is considered, the evaluated neighbours are texts, and vice-versa. We refer to \textbf{DCM-Binned} and \textbf{DCM-Continuous} as the diachronic model with binned (section~\ref{subsec:binned_diachronic}) and continuous (section~\ref{subsec:continuous_structure}) structure.

\begin{figure*}[t]
    \centering
    \includegraphics[width=1.0\linewidth]{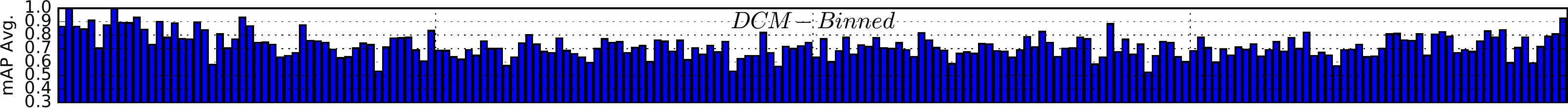}
     \includegraphics[width=1.0\linewidth]{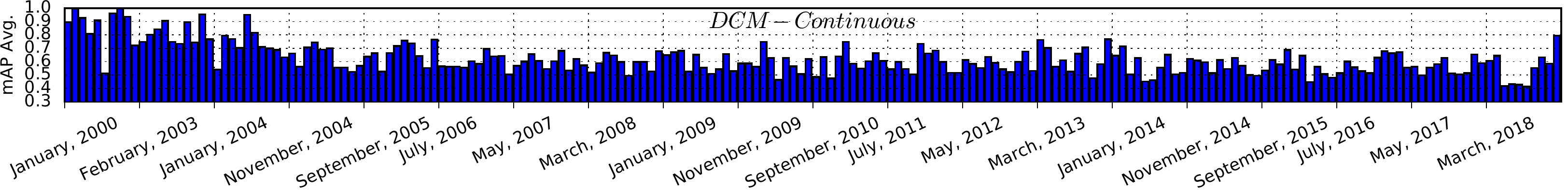}
     \vspace{-20pt}
    \caption{Temporally bounded cross-modal results (mAP) of (a) DCM-Binned (aligned) and (b) DCM-Continuous.}
  \label{fig:plots_map_per_bin}
\end{figure*}

\subsection{Diachronic Semantic Alignment}
In this section we evaluate the capability of both DCM-Binned and DCM-Continuous to capture and model diachronic data behaviours.
First it is important to assess the semantic alignment over time of the diachronic space. This corresponds to assessing if the obtained diachronic embedding space is capable of relating embeddings of images and texts of instant $ts^i$, with instances that occurred in distinct time instants $ts^j \in TS$. To accomplish this, we evaluate the semantic alignment quality of the diachronic space by designing two complementary tasks.

\vspace{5pt}
\noindent\textbf{Coarse Semantic  Alignment}
One operation permitted by a diachronic embedding consists of understanding, which instances are related to an image or text, from the set of all instances, spanning across all possible time instants. Hence, the \textit{coarse} designation. An instance $d^i$ is expected to be semantically correlated with instances not only on the $d^i$ time instant, but also on other instants (\eg recurrent events). To capture such behaviour, diachronic embedding models are required to correctly align embeddings over different time instants, such that semantic correlations are preserved.
To do this, an image or text, should be embedded on the instant $ts^i$ corresponding to its timestamp $ts^i$. Its neighbourhood in embedding space, can then be analysed by comparing the similarities of the each $d^i$ against all projected instances, on their corresponding time instant.

This is evaluated by projecting each image/text of an instance $d^i$ from the test set, in the time instant $ts^i$ (its timestamp) and evaluating the semantic similarity of its text/image neighbours, respectively. We use $mAP$, computed over the whole test set, to evaluate the similarity of the neighbours using semantic category information. The top part of Table~\ref{table:embeddings_semantic_alignment} shows the results of this experiment. The DCM-Continuous significantly outperforms DCM-Binned, revealing superior alignment capabilities. This is justified by the binning procedure, which despite the alignment procedure, due to the stochastic nature of neural networks models (\eg different initializations), leads to different organisation of data after convergence, despite its semantic correlations.

\begin{table}[t]
\caption{Diachronic Semantic Alignment.}
\label{table:embeddings_semantic_alignment}
\centering
    \begin{adjustbox}{center}
        \resizebox{0.75\columnwidth}{!}{%
\begin{tabular}{lccc}
\toprule
\multicolumn{4}{c}{\textbf{Coarse Semantic Alignment}} \\
\hline
 \multirow{1}{*}{\textbf{Methods} ($mAP$)} & $I\mapsto T$ & $T\mapsto I$ & Avg. \\
\hline
DCM-Binned (w/ Align~\cite{hamilton2016diachronic}) & 0.203 & 0.197 &  0.200\\
DCM-Continuous & 0.370 & 0.348 & 0.359\\
\hline
\multicolumn{4}{c}{\textbf{Local Semantic Alignment}} \\
\hline
 \multirow{1}{*}{\textbf{Methods} ($mAP@10$)} & $I\mapsto T$ & $T\mapsto I$ & Avg. \\
\hline
DCM-Binned (w/ Align~\cite{hamilton2016diachronic}) & 0.078 & 0.086 &  0.082\\
DCM-Continuous & 0.313 & 0.330 & 0.322\\
\bottomrule
\end{tabular}}%
\end{adjustbox}
\vspace{-18pt}
\end{table}%

\vspace{5pt}
\noindent\textbf{Local Semantic Alignment.}
In this experiment we project instances $d^i$ onto all possible timestamps $ts^j\in TS$ and assess how each projection $\vec{e}^{i, ts^j}_{*}$ relates to instances of that temporal neighbourhood (bin in our case). This operation is possible due to DCM's preservation of local alignment (\wrt to time). Namely, semantically similar instances should be close in the embedding space when projected into the same time instant $ts^j$

We sampled 50 query instances from each category, from the test set, and project each instance $d^i$ into each timestamp $ts^j \in TS$.
Then, for each time instant $ts^j$, we consider only the neighbours of the embedding of $d^i$ on that instant, \ie only embeddings $\vec{e}^{i, ts^j}_{*}$ of images and texts from $ts^j$ instant are considered.
Then we evaluate if the top-10 closest neighbours on each time instant are semantically similar (\ie belong to the same category) using  $mAP@10$. Table~\ref{table:embeddings_semantic_alignment} shows that DCM-Continuous clearly outperforms DCM-Binned. Again, this is a result of the bad alignment of the binned approach, from which DCM-Continuous is able to overcome.

\begin{figure}[t]
    \centering
    \includegraphics[width=0.26\linewidth]{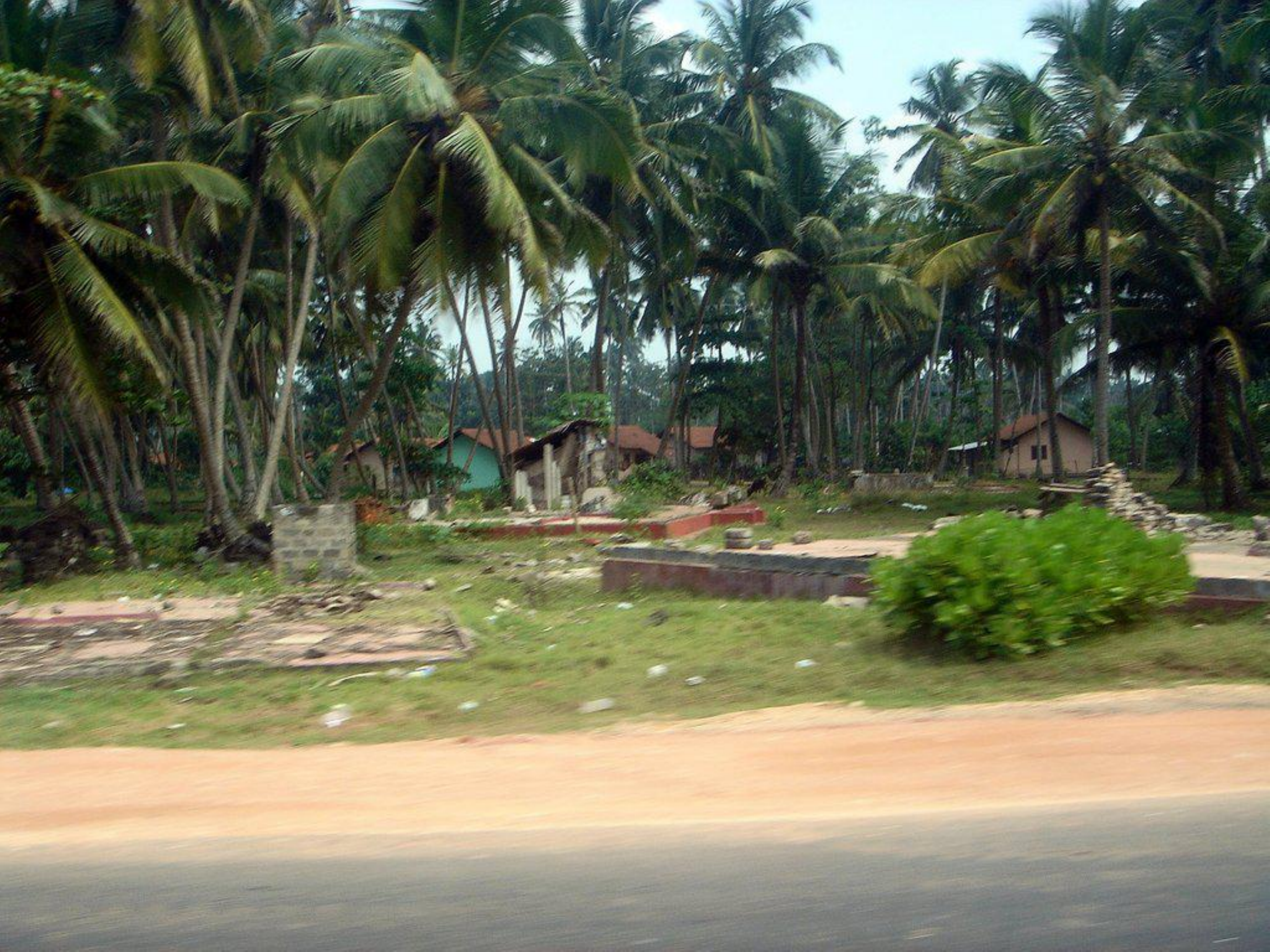}
     \includegraphics[width=0.73\linewidth]{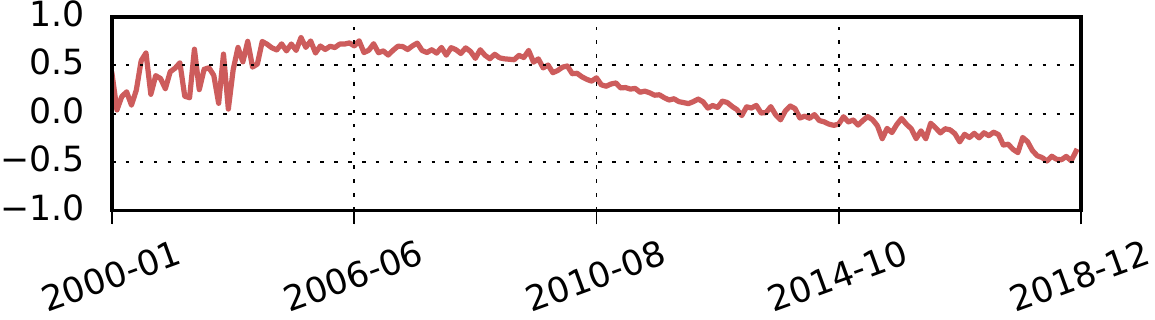}

    \includegraphics[width=0.26\linewidth]{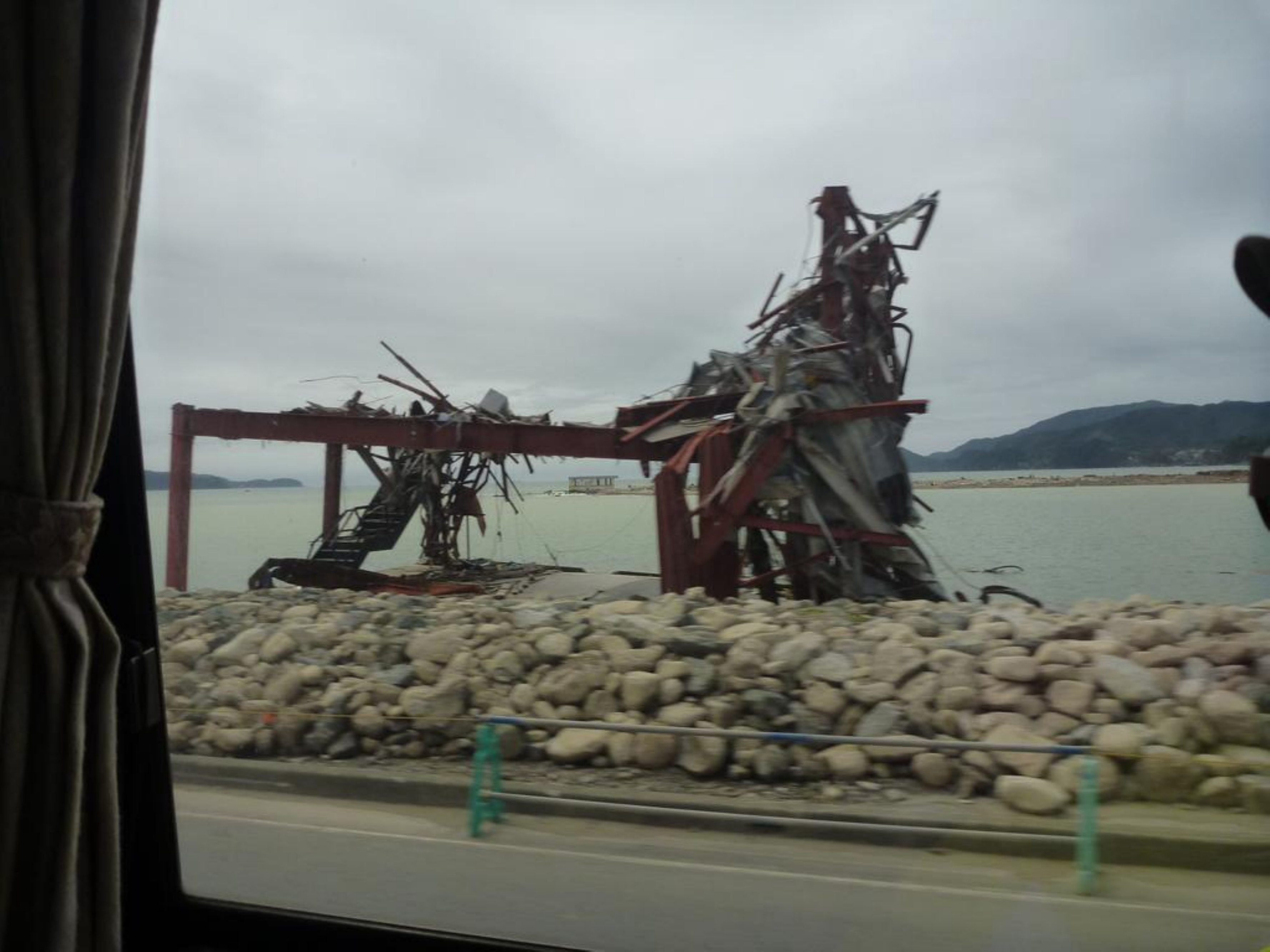}
     \includegraphics[width=0.73\linewidth]{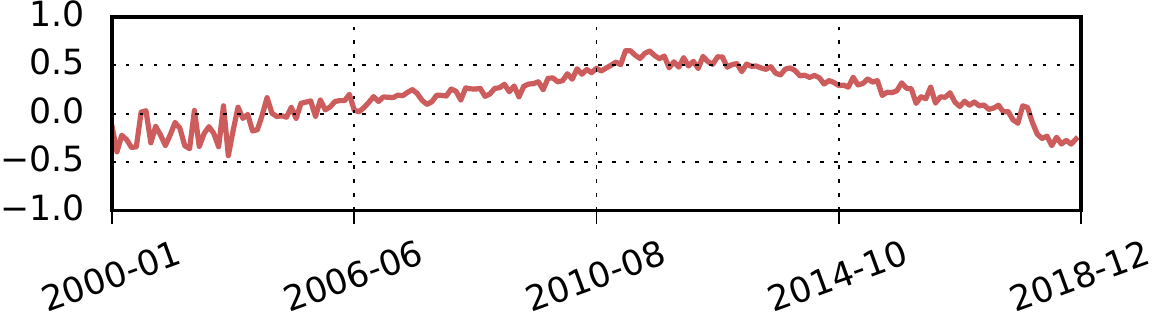}

     \includegraphics[width=0.26\linewidth]{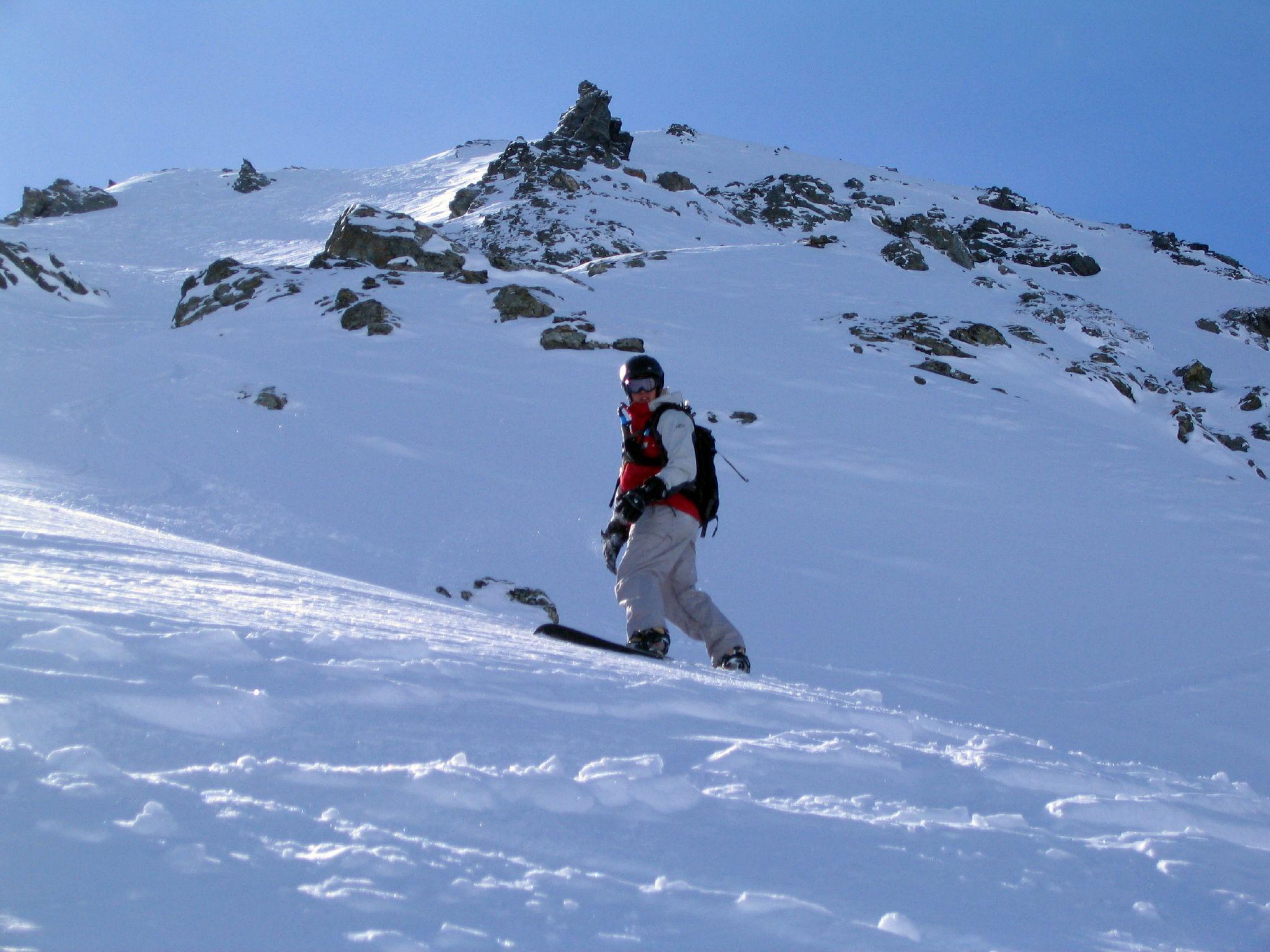}
     \includegraphics[width=0.73\linewidth]{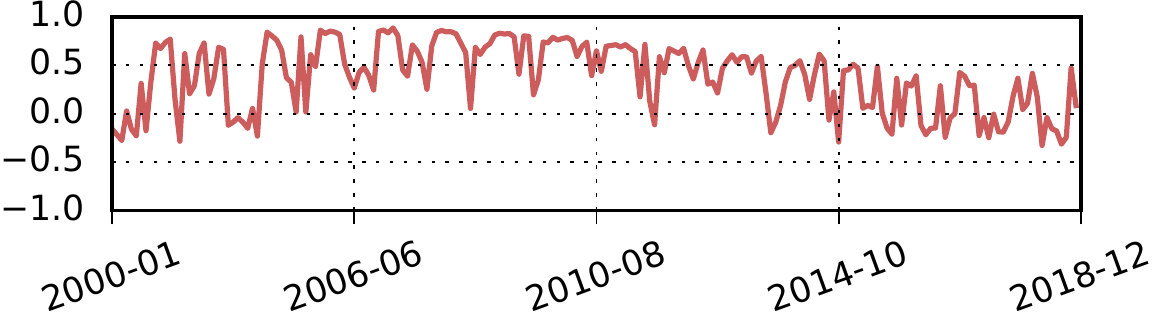}

    \includegraphics[width=0.26\linewidth]{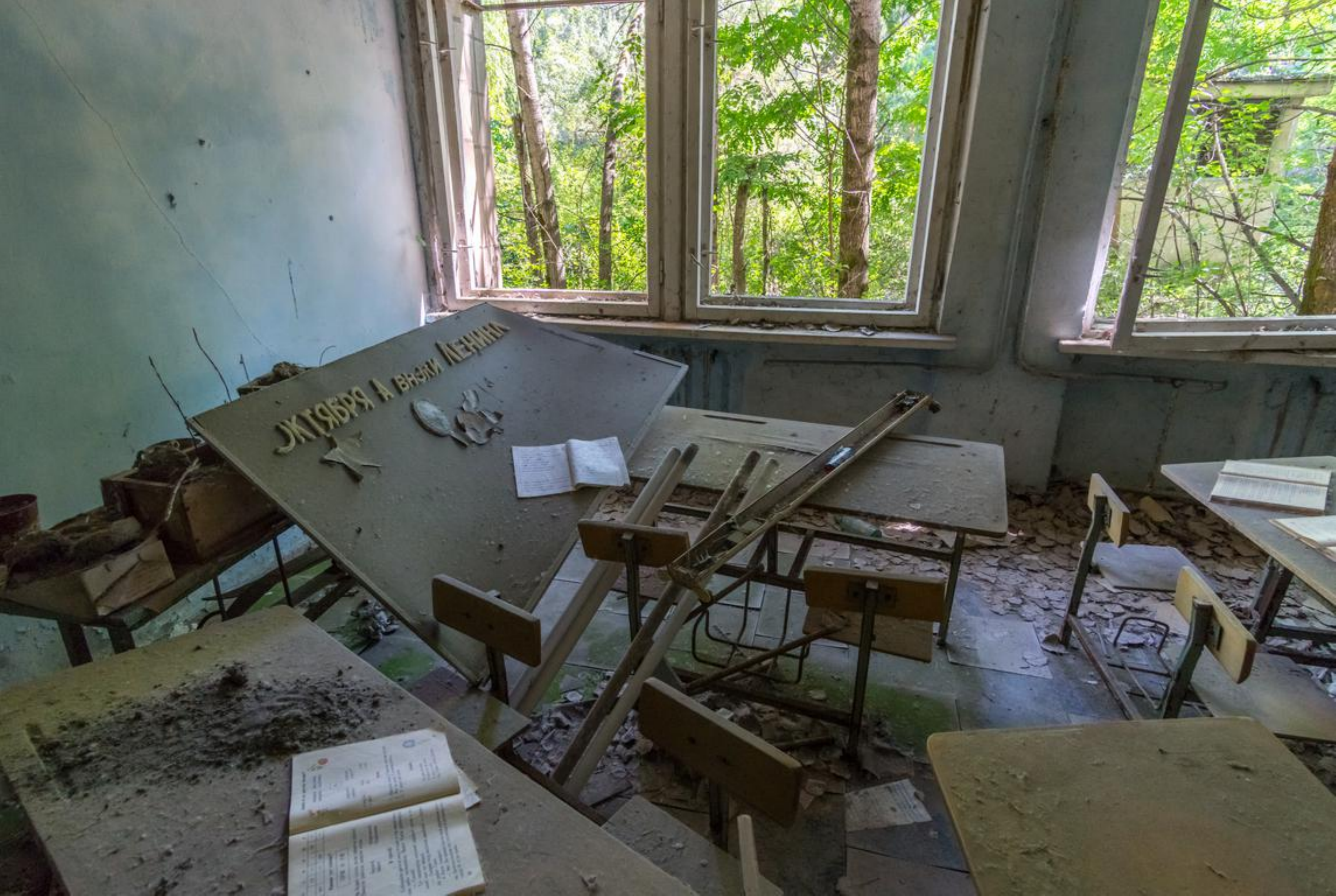}
     \includegraphics[width=0.73\linewidth]{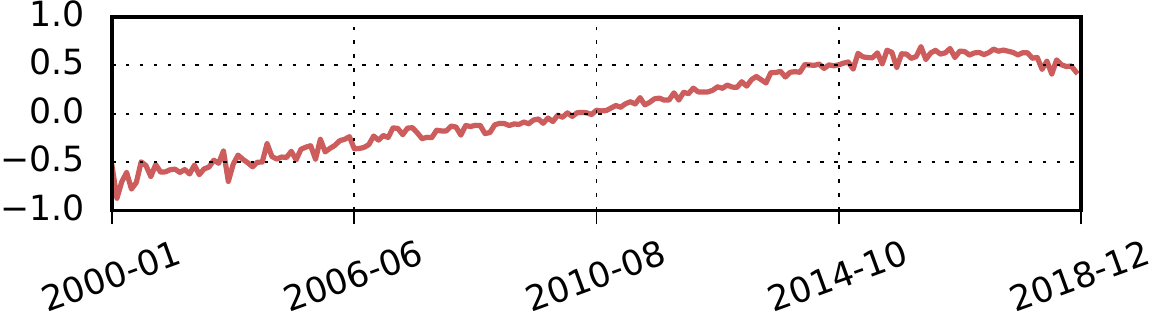}
    \caption{Semantic dispersion over time analysis for four sampled images.}
  \label{fig:dispersion}
  \vspace{-23pt}
\end{figure}

\begin{figure*}[!t]
    \centering
    \includegraphics[width=1.0\linewidth]{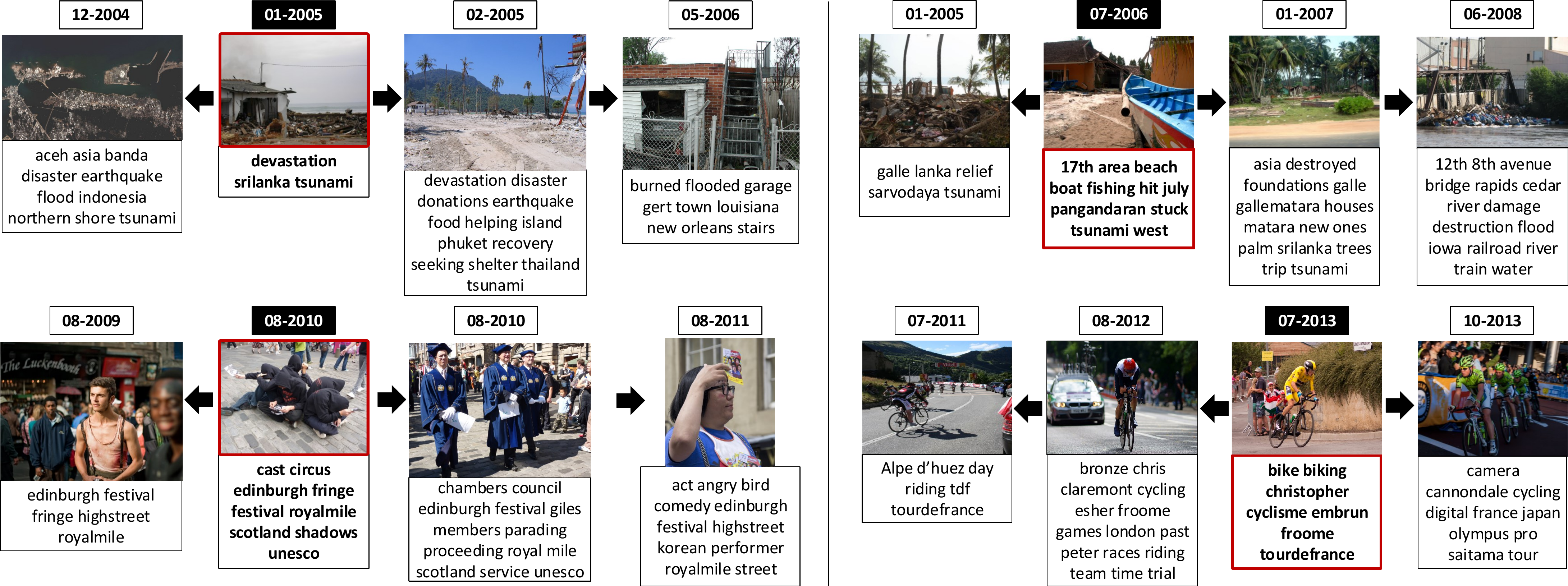}
    \caption{Evolution over time for 4 query examples (query timestamps are black-filled). Instances were retrieved from before and after the query timestamp. On the left, image queries were used to retrieve documents through their text. On the right, text queries were used to retrieve documents through their images.}
  \label{fig:instances_evolution_img2txt}
\end{figure*}

\subsection{Temporally-bounded Semantics}
\label{subsec:temporally_bounded}
In this section we evaluate the semantic organisation of the embedding space in temporally-bounded data. To evaluate the correctness of the embeddings neighbourhood at each instant, the test set is binned and each bin is evaluated individually.
Following cross-modal learning works~\cite{Rasiwasia:2010:NAC:1873951.1873987, Wang2016ACS, Feng:2014:CRC:2647868.2654902,Wang:2017:ACR:3123266.3123326, 7298966, 8013822}, we evaluate each method on the tasks of \emph{Image-to-Text}  and \emph{Text-to-Image} retrieval, using mean Average Precision $mAP$ for \emph{all the  retrieved results}. Figure~\ref{fig:plots_map_per_bin} shows the plots of DCM variants, with the $mAP$ results per month. A static cross-modal embedding baseline, corresponding to our model without the temporal layer and with $\mathcal{L}_{intra}(\cdot)=0$, was trained in all data ignoring time and obtained a $mAP$ of 0.639. The DCM-Binned baseline and the DCM-continuous methods obtained a $mAP$ of 0.724 and 0.623, respectively. We further evaluated against TempXNet~\cite{Semedo:2018:TCR:3240508.3240665}, which uses temporal correlations to learn a static embedding space, obtaining a $mAP$ of 0.645. All DCM variants have shown to be on par with state-of-the-art approaches, by obtaining scores above $0.60$ $mAP$ points. This experiment confirms our hypothesis that semantic cross-modal correlations change over time. DCM-Binned outperformed all the other methods due to its temporal locality bias: for each bin (month) a static model is trained \emph{solely} on data from that bin, thus modelling the local cross-modal correlations independently, and without influences from correlations of the remaining bins. However, it lacks the advantages of a fully diachronic model.

\subsection{Media Time-Period Inference}

We now evaluate the enforcement of properties that define the neighbourhood of each projected instance, based on a given time window (Properties 1 and 2). Accordingly, we evaluate the neighbours of each image/text $d^i$, by computing a \textit{temporally bounded mAP} ($mAP@50$): a neighbour is considered relevant if it belongs to the same category and its timestamp is within a time-window of size $w$, \wrt to the timestamp $ts^i$ of $d^i$. The results are shown in table~\ref{table:time_period_inference_results}. It can be observed that DCM-Continuous significantly outperforms the two compared baselines in defining neighbourhoods that respect property 1, \ie that instances from the same category that are close in time (time window $w$), lye close together.
Section~\ref{subsec:temporally_bounded} and Figure~\ref{fig:plots_map_per_bin} show that static cross-modal methods need to be artificially fed with the relevant time period to achieve comparable results. In contrast, this result experimentally demonstrates that DCM-Continuous does not need to receive this outside information and can indeed infer the relevant time periods on its own.

\begin{table}[t]
\caption{Time-Period Inference Results.}
\vspace{-10pt}
\label{table:time_period_inference_results}
\centering
\begin{tabular}{lccc}
\toprule
 \multirow{1}{*}{\textbf{Methods} ($mAP@50$)} & $I\mapsto T$ & $T\mapsto I$ & Avg. \\
\hline
Static cross-modal & 0.048 & 0.059 &  0.054\\
TempXNet~\cite{Semedo:2018:TCR:3240508.3240665} & 0.052 & 0.070 & 0.061\\
DCM-Continuous & \textbf{0.126} & \textbf{0.144} & \textbf{0.135}\\
\bottomrule
\end{tabular}%
\vspace{-15pt}
\end{table}%

\subsection{Semantic Dispersion over Time}
Previously, we confirmed that diachronic embedding spaces are semantically aligned and that semantic correlations do change over time. Hence, in this section we examine the semantic dispersion of multimodal data over time. Given an image or a text, we expect that its correlations with other instances, over time, will evolve. The evolution pattern is expected to be grounded on the temporal characteristics (\eg peak based, recurrent event, etc.) of the topic of each instance.

To assess this, we consider a set of target instances $d^i$, in which semantic evolution will be evidenced by semantic dispersion changes, over each instant $ts \in TS$. Namely, for an instance $d^i$ with timestamp $ts^i$, given its embedding on instant $ts^i$, we define semantic dispersion as the variation of the similarity between $d^i$ embedding and its closest neighbour instances. In practice, we sample instances $d^i$ and project them in the time instant $ts^i$ corresponding to its timestamp, using DCM-Continuous, obtaining the embedding $\vec{e}^{i,ts^i}$. Then, for each instant $ts\in TS$, we compute the semantic dispersion of the top $K$ neighbours of $\vec{e}^{i,ts^i}$. Semantic dispersion on an instant $ts$ is defined as the average of the cosine similarities between $\vec{e}^{i,ts^i}$ and each of the $K$ neighbours. We set $K=5$.

Figure~\ref{fig:dispersion} shows the results of this experiment for four different images. The first two images belong to the \textit{tsunami} category: the first corresponds to the Indonesia series of tsunamis in 2007, and the second to the tsunami in Japan, 2012. It can be seen that maximal similarity is achieved around the dates of the corresponding tsunamis. For the first image, after the peak around 2006, similarity decreases gradually in future instants. Despite the tsunami of 2012, its similarity with content from that tsunami is low. This evidences that DCM-Continuous effectively delivers Property 4 (section~\ref{subsec:properties}). The third image, \textit{snowboarding}, shows a recurrent evolution of semantic correlations over time, being more stable over the years. The last image, taken in August 2018, depicts an abandoned place due to the \emph{nuclear disaster} of Chernobyl (pictures of contaminated areas became possible with the advent of drones and robots). Semantic similarity over time in this image gradually increases until 2015, and then stabilises until 2018. This experiment shows that diachronic embeddings obtained with DCM-Continuous encode cross-modal interactions evolution, enabling the understanding of multimodal correlations over time.

\subsection{Cross-Modal Evolution}
The cross-modal diachronic embedding model enables novel ways of exploring multimodal instances. One example is the analysis of the correlations evolution of an image or text, Figure~\ref{fig:instances_evolution_img2txt}, along the years, which is encoded in its embedding trajectory. Such operation enables one to understand the correlations shift along the years.

To perform this experiment, we sample a set of images and texts, and project them on their corresponding timestamp instant. Then, to avoid inspecting all time instants, we restrict the number of instants to the top-20 bin, \ie the bins with the closest image/text, to a target text/image, respectively, based on cosine similarity. For illustration purposes, we show images/texts from 4 different instants on Figure~\ref{fig:instances_evolution_img2txt}. Queries are marked with the black-filled timestamps and instances were retrieved from before and after this timestamp.

The inspection of the evolution timeline (obtained with DCM-Continuous) lets us interpret the trajectory of the sampled images and texts, at particular time instants.
For example, on the top-left case, the query was an image of the 2004 tsunami in Indonesia. DCM-Continuous retrieved texts of other tsunamis and floods that occurred over the years.
On the right-bottom example of Figure~\ref{fig:instances_evolution_img2txt} the query was a textual description of a picture with cyclist Christopher Froome. With this textual query and corresponding timestamp, DCM-Continuous was able to illustrate the evolution of \textit{cyclism} topic over time.

\section{Conclusions}
This paper introduced the first diachronic cross-modal embedding, enabling novel interpretations of cross-modal semantic shifts over time. The key novelties of the proposed diachronic model are:
\begin{itemize}
    \item The neural architecture implements a \textit{temporal structuring layer} that is shared across the two projection functions.
    \item A novel \textit{joint diachronic ranking loss} function to control temporal structuring of data in the embedding space.
\end{itemize}

Moreover, experiments, on a 20 year span dataset, illustrated the semantic evolution and temporal flexibility of the model. The key take away messages are:
\begin{itemize}
    \item \textit{Cross-modal semantic evolution} is captured by the model allowing the inspection of temporal multimodal information;
    \item \textit{Time is handled in a flexible manner}, \ie projections are timestamped, data is organised temporally, thus supporting several diachronical operations.
\end{itemize}

As future work, we plan to investigate the geometry of the embedding space when using other temporal distributions to capture complex multimodal interactions over time.

\begin{acks}
This work has been partially funded by the \grantsponsor{CMUP}{CMU Portugal}{} research project GoLocal Ref. \grantnum{CMUP}{CMUP-ERI/TIC/0046/2014}, by the \grantsponsor{EU H2020}{H2020 ICT}{} project COGNITUS with the grant agreement n\textsuperscript{o} \grantnum{EU H2020}{687605} and by the \grantsponsor{FCT}{FCT}{} project NOVA LINCS Ref. \grantnum{FCT}{UID/CEC/04516/2019}. We also gratefully acknowledge the support of NVIDIA Corporation with the donation of the GPUs used for this research.
\end{acks}

\balance
\bibliographystyle{ACM-Reference-Format}
\bibliography{02_bibliography}

\end{document}